\documentstyle[12pt,epsf]{article}
\def \beq{\begin{equation}}
\def \eeq{\end{equation}}
\def \s{\sqrt{2}}
\def \st{\sqrt{3}}
\def \sx{\sqrt{6}}
\begin{document}
\rightline{TECHNION-PH-95-27}
\rightline{EFI-95-60}
\rightline{hep-ph/9509428}
\rightline{September 1995}
\bigskip
\bigskip
\centerline{{\bf AMPLITUDE RELATIONS FOR $B$ DECAYS INVOLVING $\eta$ and
$\eta'$}\footnote{To be submitted to Physics Letters B}}
\bigskip
\centerline{\it Amol S. Dighe}
\centerline{\it Enrico Fermi Institute and Department of Physics}
\centerline{\it University of Chicago, Chicago, IL 60637}
\medskip
\centerline{\it Michael Gronau}
\centerline{\it Department of Physics}
\centerline{\it Technion -- Israel Institute of Technology, Haifa 32000,
Israel}
\medskip
\centerline{and}
\medskip
\centerline{\it Jonathan L. Rosner}
\centerline{\it Enrico Fermi Institute and Department of Physics}
\centerline{\it University of Chicago, Chicago, IL 60637}
\bigskip
\centerline{\bf ABSTRACT}
\medskip
\begin{quote}

A class of amplitude relations for decays of $B$ mesons is discussed. Processes
involving $\eta$ and $\eta'$ in the final state are shown to provide useful
information about weak phases in some cases even in the presence of
octet-singlet mixing in these states.  Some of the relations are unaffected by
first-order SU(3) breaking.
\end{quote}
\bigskip

The decays of $B$ mesons can provide unique insights into the details of the
weak interactions.  Among the most eagerly sought of these is a confirmation
that phases in the elements of the Cabibbo-Kobayashi-Maskawa (CKM) matrix
\cite{CKM} are responsible for the observed violation of CP invariance
\cite{CCFT} in decays of neutral kaons.

In the present report we describe a class of $B$ decays which can shed light on
CKM phases through the construction of relations satisfied by complex
decay amplitudes. Previous results \cite{BPP,PRL,PLB} involving the decays $B
\to \pi \pi$, $B \to \pi K$, and $B \to K \bar K$ have indicated the
possibility of isolating CKM phases from the effects of strong interactions
using such relations, but the constructions were complicated by the
presence of higher-order electroweak (``penguin'') contributions
\cite{RF,DH,DHT,EWP}. The inclusion of decays involving $\eta$ mesons
\cite{Desh} provides enough information to avoid this complication as long as
$\eta$ is regarded as a flavor octet member.  Our discussion, involving $\eta$
and $\eta'$ mesons, takes account of the observed octet-singlet mixing in the
$\eta$ and $\eta'$ \cite{GK}.  A more complete discussion is presented in
Ref.~\cite{Dighe}.

\begin{figure}
% \vspace{1.9in}
\centerline{\epsfysize = 1.9in \epsffile{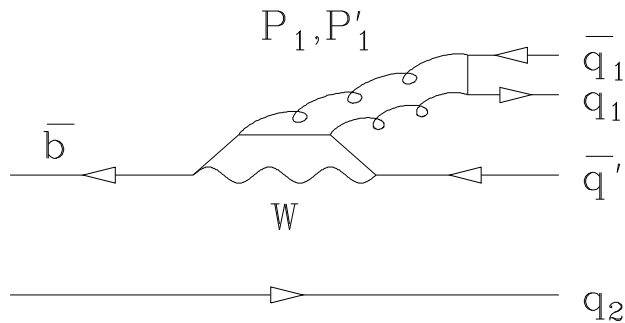}}
\caption{Graph contributing to $B$ decays involving one or two flavor SU(3)
singlet pseudoscalar mesons.  The coiled lines denote a color singlet exchange
due to two or more gluons.  Here $q_1,~q_2 = u,d,s$ while $q' = d,s$.}
\end{figure}

Our method employs flavor SU(3) symmetry \cite{DZ,SW,Chau,SilWo} in order to
relate decays involving pairs of light pseudoscalar mesons in the final
state. We present results for the case, very close to experiment
\cite{GK,Chau}, in which $\eta = (s \bar s - u \bar u - d \bar d)/\st$ and
$\eta' = (2 s \bar s + u \bar u + d \bar d)/\sx$.  We retain only those
amplitudes in which the spectator quark (the light quark accompanying the $b$
in the initial meson) does not enter into the decay Hamiltonian.  These
amplitudes are expected to be the dominant ones; others are expected to be
suppressed by a power of $f_B/m_B$ or $f_{B_s}/m_{B_s}$, where $f_B \simeq 180$
MeV or $f_{B_s} \simeq 225$ MeV are decay constants.  Most of our results are
unaffected by first-order SU(3) breaking, for which a partial set of tests has
been presented elsewhere \cite{SU3}.

In the present approximation \cite{Dighe,Quad} there are four independent
amplitudes:  a ``tree'' contribution $t$, a ``color-suppressed'' contribution
$c$, a ``penguin'' contribution $p$, and a ``singlet penguin'' contribution
$p_1$, in which a color-singlet $q \bar q$ pair produced by two or more gluons
(see Fig.~1) or by a $Z$ or $\gamma$ forms an SU(3) singlet state . These
amplitudes contain both the leading-order and electroweak penguin
contributions:
$$
t \equiv T + (c_u - c_d) P_{EW}^C~~,~~~c \equiv C + (c_u - c_d) P_{EW}~~~,
$$
\beq \label{eqn:dict}
p \equiv P + c_d P_{EW}^C~~,~~~p_1 \equiv P_1 + c_d P_{EW}~~~,
\eeq
where the capital letters denote the leading-order contributions (as defined,
for example, in Ref.~\cite{BPP}), and $P_{EW}$ and $P_{EW}^C$ are color-favored
and color-suppressed electroweak penguin amplitudes, as defined in
Ref.~\cite{EWP}. Here $c_u$ and $c_d$ depend on the structure of the
electroweak penguin; amplitudes were redefined in Ref.~\cite{EWP} such that
$c_u = 2/3$ and $c_d = -1/3$.  As shown earlier \cite{EWP}, the electroweak
penguin contributions may be incorporated without affecting SU(3) analyses as
long as one does not attempt to relate $\Delta S = 0$ and $|\Delta S| = 1$
processes. In what follows we shall denote the $\Delta S = 0$ processes by
unprimed quantities and the $|\Delta S| = 1$ processes by primed quantities.

The results for decays involving $\eta$ and $\eta'$ are shown in Tables 1
and 2.  Decompositions of amplitudes for $\pi \pi$, $\pi K$, and $K \bar K$
decays in terms of the three contributions $t(t'),~c(c')$, and $p(p')$
have been given elsewhere \cite{BPP}. Our notation \cite{Quad} is
equivalent to that of Ref.~\cite{Dighe}.

\renewcommand{\arraystretch}{1.2}
\begin{table}
\caption{Decomposition of $B \to PP$ amplitudes involving physical $\eta$ and
$\eta'$ (as defined in the text) for $\Delta C = \Delta S = 0$ transitions in
terms of graphical contributions.}
\begin{center}
\begin{tabular}{l l c c c c} \hline
         & Final        &    $t$   &    $c$   &    $p$   &    $p_1$ \\
         & state        &          &          &          &          \\ \hline
$B^+\to$ & $\pi^+\eta$  & $-1/\st$ & $-1/\st$ & $-2/\st$ & $-1/\st$  \\
         & $\pi^+\eta'$ &  $1/\sx$ &  $1/\sx$ &  $2/\sx$ & $2\sqrt{2/3}$ \\
\hline
$B^0\to$ & $\pi^0\eta$  &     0    &     0    & $-2/\sx$ & $-1/\sx$  \\
         & $\pi^0\eta'$ &     0    &     0    &  $1/\st$ & $2/\st$   \\
         & $\eta\eta$   &     0    &  $\s/3$  &  $\s/3$  & $\s/3$   \\
         & $\eta\eta'$  &     0    & $-\s/3$  & $-\s/3$  & $-5/3\s$ \\
         & $\eta'\eta'$ &     0    &  $\s/6$  &  $\s/6$  & $2\s/3$  \\ \hline
$B_s\to$ & $\eta \bar K^0$ &   0   & $-1/\st$ &     0    & $-1/\st$  \\
         & $\eta' \bar K^0$ &  0   &  $1/\sx$ &  $3/\sx$ & $2\sqrt{2/3}$ \\
\hline
\end{tabular}
\end{center}
\end{table}

\begin{table}
\caption{Decomposition of $B \to PP$ amplitudes for $\Delta C = 0,~ |\Delta S|
= 1$ transitions involving physical $\eta$ and $\eta'$ (as defined in the text)
in terms of graphical contributions.}
\begin{center}
\begin{tabular}{l l c c c c} \hline
         & Final       &     $t'$  &     $c'$  &  $p'$   &   $p_1'$  \\
         & state       &           &           &         &           \\ \hline
$B^+\to$ & $\eta K^+$  &  $-1/\st$ &  $-1/\st$ &   0     &  $-1/\st$  \\
         & $\eta' K^+$ &   $1/\sx$ &   $1/\sx$ & $3/\sx$ & $2\sqrt{2/3}$ \\
\hline
$B^0\to$ & $\eta K^0$  &      0    &  $-1/\st$ &   0     &  $-1/\st$  \\
         & $\eta' K^0$ &      0    &   $1/\sx$ & $3/\sx$ & $2\sqrt{2/3}$ \\
\hline
$B_s\to$ & $\pi^0 \eta$ &     0    &  $-1/\sx$ &   0     &    0      \\
         & $\pi^0 \eta'$ &    0    &  $-1/\st$ &   0     &    0      \\
         & $\eta\eta$  &      0    &  $-\s/3$  & $\s/3$  &  $-\s/3$  \\
         & $\eta\eta'$ &      0    &  $-1/3\s$ & $2\s/3$ &   $\s/3$  \\
         & $\eta'\eta'$ &     0    &   $\s/3$  & $2\s/3$ &  $4\s/3$  \\ \hline
\end{tabular}
\end{center}
\end{table}

Several amplitude sum rules involving $\eta$ and $\eta'$ are implied by the
Tables. A more complete list (including some $B_s$ decays omitted here) may be
found in Ref.~\cite{Dighe}.  We label those which are unaffected by first-order
SU(3) breaking \cite{SU3} by an asterisk.
\bigskip

\leftline{\it $\Delta S = 0$ decays:}
\medskip

*1.  A relation involving only pions and $\eta$'s is
$$
\s A(B^+ \to \pi^+ \pi^0) - \st A(B^+ \to \pi^+ \eta) + \sx A(B^0 \to \pi^0
\eta)
$$
\beq \label{eqn:pie}
= -(t + c) + (t + c + 2p + p_1) - (2p + p_1) = 0~~~.
\eeq
Since the $B^0 \to \pi^0 \eta$ process involves only $p$ and $p_1$ (including
electroweak penguin contributions), its weak phase is approximately Arg $V_{td}
= - \beta$, if we neglect corrections \cite{BF} due to quarks other than the
top quark. The weak phase of $t+c$ is approximately Arg $V_{ub}^* = \gamma$, if
one neglects electroweak penguin contributions which should be small in this
case \cite{EWP}. The relative weak phase of the two amplitudes is then $\gamma
+ \beta = \pi - \alpha$, where $\alpha,~\beta$, and $\gamma$ are the three
angles of the unitarity triangle.  By comparing the triangle (\ref{eqn:pie})
with the corresponding one for the charge-conjugate processes, one can then
measure not only $\alpha$, but also the strong phase difference between the
$t+c$ and $2p + p_1$ terms.

The charged $B$ decays are ``self-tagging.''  The $B^0 \to \pi^0 \eta$ and
$\bar B^0 \to \pi^0 \eta$ rates are expected to be equal since the amplitudes
$p$ and $p_1$ should have equal weak phases.  Thus the only shortcoming of this
triangle relation is the expected smallness of the $B^0 \to \pi^0 \eta$ decay
rate, which we estimate (using $|p| = {\cal O}(1/5)|t|$ and neglecting the
unknown $|p_1|$ contribution) to be less than $10^{-6}$.  The amplitude
triangles should have two long sides and a short one.
\medskip

*2.  A relation involving only $\pi^0$'s and $\eta$'s is
$$
\s A(B^0 \to \pi^0 \pi^0) + \sx A(B^0 \to \pi^0 \eta) + \frac{3}{\s}
A(B^0 \to \eta \eta)
$$
\beq \label{eqn:neuts}
= (p-c) - (2p + p_1) + (c + p + p_1) = 0~~~.
\eeq
In this case, in contrast to the previous one, we have to identify the flavor
at time of production of the decaying neutral $B$, since the rates for the
$\pi^0 \pi^0$ and $\eta \eta$ decays are not necessarily expected to be equal
for $B^0$ and $\bar B^0$ decays.  Since all the amplitudes are either penguins
or color-suppressed, we expect all sides of the amplitude triangles to be
small, corresponding to branching ratios of $10^{-6}$ or less.
\medskip

*3.  In the absence of the $p_1$ contribution, we would have
\beq
A(B^+ \to \pi^+ \eta) = -\s A(B^+ \to \pi^+ \eta') = -(t+c+p)/\st~~~,
\eeq
since the $t$, $c$, and $p$ contributions to these processes probe only the
nonstrange quark content of the $\eta$ and $\eta'$.  Similarly, if $p_1$ is
neglected, we have
\beq
A(B^0 \to \pi^0 \eta) = -\s A(B^0 \to \pi^0 \eta') = -2p/\sx~~~.
\eeq
The degree of violation of these relations thus can be regarded as one measure
of the importance of the $p_1$ term.  The relation for neutral $B$ decays is
probably more sensitive to $p_1$ since it lacks the dominant $t$ contribution.
The $p_1$ term receives contributions from both $P_1$, illustrated in Fig.~1,
in which the flavor singlet $q_1 \bar q_1$ pair is connected to the rest of the
diagram with at least two gluons, and the electroweak penguin term $P_{EW}$, as
indicated in the last of Eqs.~(\ref{eqn:dict}).  It is quite likely that each
of these is smaller than the $p$ term, which in turn is likely to be ${\cal O}
(1/5)$ of the dominant $t$ term.
\medskip

4.  The following amplitudes are expected to have approximately the same
magnitudes for the $B$ and $\bar B$ decay processes, since they have a common
weak phase in the limit of $t$ dominance of $p$ and $p_1$:
$$
3 A(B^+ \to K^+ \bar K^0) + 2 \sx A(B^0 \to \pi^0 \eta) + \st A(B^0 \to
\pi^0 \eta')
$$
\beq \label{eqn:allp}
= 3p - 2(2p + p_1) + (p + 2p_1) = 0~~~.
\eeq
Thus, in this approximation,
one does not expect to see CP violation in any of the rates for these
processes, and flavor tagging is unnecessary.  The shape of the amplitude
triangle provides information on the relative magnitudes and phases of $p$
and $p_1$.  The above relation is affected by first-order SU(3) breaking
because it involves creation from the vacuum of both nonstrange and
strange quark pairs.
\medskip

*5.  There is an amplitude triangle involving decays with $\eta$ and $\eta'$:
$$
A(B^0 \to \eta \eta) + 2 A(B^0 \to \eta \eta') + 2 A(B^0 \to \eta' \eta')
$$
\beq
= (\s/3)(c+p+p_1) -(\s/3)(2c + 2p + 5p_1) + (\s/3)(c + p + 4p_1) = 0~~~,
\eeq
from which we can learn the relative weak and strong phases of $c+p$ and $p_1$.
Since we learn $p_1/p$ from the previous construction, we can then learn
$c/p$.  Since electroweak penguins probably contribute in a non-negligible
manner to $c$, this information may be of limited use in determining weak
phases.
\medskip

\leftline{\it $|\Delta S| = 1$ decays:}
\medskip

1.  A quadrangle relation \cite{Quad} for charged $B$ decays generalizes the
triangle relation of Ref.~\cite{Desh}:
$$
2 A(B^+ \to \pi^+ K^0) + \s A(B^+ \to \pi^0 K^+) -
(4/\st) A(B^+ \to \eta K^+) - (\sx/3) A(B^+ \to \eta' K^+)
$$
\beq
= 2p' - (t' + c' + p') + (4/3)(t' + c' + p_1') - (1/3)(t' + c' + 3p' + 4p_1')
= 0~~~.
\eeq
It is possible to inscribe in the quadrangle a triangle composed of linear
combinations of $p'$ and $p_1'$, whose shape does not change under charge
conjugation.  Thus, by studying the processes and their charge conjugates,
one can construct rigid quadrangles (up to discrete ambiguities), from which
-- in the manner of Refs.~\cite{EWP} and \cite{Desh}, making use of the
rate for $B^+ \to \pi^+ \pi^0$ -- one can learn the weak phase $\gamma$
\cite{Quad}.  The absence of the term $p'$ in the amplitude for $B^+ \to \eta
K^+$ (which arises as a result of cancellation of nonstrange and strange quark
contributions for the particular mixture of octet and singlet assumed here)
suggests that this may be the process with the smallest branching ratio of the
four.  This relation is affected by first-order SU(3) breaking for the same
reason as in Eq.~(\ref{eqn:allp}).
\medskip

2.  A triangle relation satisfied by the neutral $B$ decays is affected by
first-order SU(3) breaking for the same reason as the previous result:
$$
3 \s A(B^0 \to \pi^0 K^0) - 4 \st A(B^0 \to \eta K^0) - \sx A(B^0 \to
\eta' K^0)
$$
\beq
= 3(p'-c') + 4(c' + p_1') - (c' + 3p' + 4p_1') = 0~~~.
\eeq
Here one learns $c'$ relative to $p'$ or $p_1'$.  This information may not be
so useful for the study of weak phases, but it can serve to check specific
models.
\medskip

*3.  The ratio of the rates for $B_s \to \pi^0 \eta$ and $B_s \to \pi^0
\eta'$ can check the relative strange quark content of the two states, since
these amplitudes both depend only on $c'$.  For the mixing adopted here, we
predict
\beq
A(B_s \to \pi^0 \eta') = \s A(B_s \to \pi^0 \eta)~~~
\eeq
The $\pi^0 \eta$ mode was suggested as one way of specifying the shape of
the isospin quadrangle in $B \to \pi K$ decays \cite{EWP}, avoiding the
problems associated with electroweak penguins \cite{DH} that arose in an
earlier construction \cite{BPP,PRL}.
\medskip

A number of additional $|\Delta S| = 1$ decays involving $B_s$ and $\eta$
or $\eta'$ are noted in Ref.~\cite{Dighe}.

To summarize, several interesting amplitude relations involving decays of $B$
mesons with $\eta$ and/or $\eta'$ in the final state can shed light on weak
phases and on relative strong phases and magnitudes of different contributions
to amplitudes.  Foremost among these are a triangle relation satisfied by the
amplitudes for $B^+ \to \pi^+ \pi^0$, $B^+ \to \pi^+ \eta$, and $B^0 \to \pi^0
\eta$, all unaffected by first-order SU(3) breaking, and a quadrangle relation
\cite{Quad} satisfied by the amplitudes for $B^+ \to \pi^+ K^0,~\pi^0 K^+, \eta
K^+$, and $\eta' K^+$.  The detection of these modes poses an interesting
instrumental problem well-matched to the capabilities of detectors at present
or planned $e^+ e^-$ colliders.

We thank H. Lipkin and S. Stone for fruitful discussions, and the Aspen Center
for Physics for a congenial atmosphere in which the main part of this
collaboration was carried out. M. G. and J. L. R. wish to acknowledge the
respective hospitalities of the SLAC and Fermilab theory groups during parts of
this investigation. This work was supported in part by the United States --
Israel Binational Science Foundation under Research Grant Agreement 94-00253/1,
by the Fund for Promotion of Research at the Technion, and by the United States
Department of Energy under Contract No. DE FG02 90ER40560.
\bigskip

% Journal and other miscellaneous abbreviations for references
% Phys. Lett. B style
\def \ajp#1#2#3{Am.~J.~Phys.~{\bf#1} (#3) #2}
\def \apny#1#2#3{Ann.~Phys.~(N.Y.) {\bf#1} (#3) #2}
\def \app#1#2#3{Acta Phys.~Polonica {\bf#1} (#3) #2 }
\def \arnps#1#2#3{Ann.~Rev.~Nucl.~Part.~Sci.~{\bf#1} (#3) #2}
\def \cmp#1#2#3{Commun.~Math.~Phys.~{\bf#1} (#3) #2}
\def \cmts#1#2#3{Comments on Nucl.~Part.~Phys.~{\bf#1} (#3) #2}
\def \cn{Collaboration}
\def \corn93{{\it Lepton and Photon Interactions:  XVI International Symposium,
Ithaca, NY August 1993}, AIP Conference Proceedings No.~302, ed.~by P. Drell
and D. Rubin (AIP, New York, 1994)}
\def \cp89{{\it CP Violation,} edited by C. Jarlskog (World Scientific,
Singapore, 1989)}
\def \dpff{{\it The Fermilab Meeting -- DPF 92} (7th Meeting of the American
Physical Society Division of Particles and Fields), 10--14 November 1992,
ed. by C. H. Albright \ite~(World Scientific, Singapore, 1993)}
\def \dpf94{DPF 94 Meeting, Albuquerque, NM, Aug.~2--6, 1994}
\def \efi{Enrico Fermi Institute Report No. EFI}
\def \el#1#2#3{Europhys.~Lett.~{\bf#1} (#3) #2}
\def \f79{{\it Proceedings of the 1979 International Symposium on Lepton and
Photon Interactions at High Energies,} Fermilab, August 23-29, 1979, ed.~by
T. B. W. Kirk and H. D. I. Abarbanel (Fermi National Accelerator Laboratory,
Batavia, IL, 1979}
\def \hb87{{\it Proceeding of the 1987 International Symposium on Lepton and
Photon Interactions at High Energies,} Hamburg, 1987, ed.~by W. Bartel
and R. R\"uckl (Nucl. Phys. B, Proc. Suppl., vol. 3) (North-Holland,
Amsterdam, 1988)}
\def \ib{{\it ibid.}~}
\def \ibj#1#2#3{~{\bf#1} (#3) #2}
\def \ichep72{{\it Proceedings of the XVI International Conference on High
Energy Physics}, Chicago and Batavia, Illinois, Sept. 6--13, 1972,
edited by J. D. Jackson, A. Roberts, and R. Donaldson (Fermilab, Batavia,
IL, 1972)}
\def \ijmpa#1#2#3{Int.~J.~Mod.~Phys.~A {\bf#1} (#3) #2}
\def \ite{{\it et al.}}
\def \jmp#1#2#3{J.~Math.~Phys.~{\bf#1} (#3) #2}
\def \jpg#1#2#3{J.~Phys.~G {\bf#1} (#3) #2}
\def \lkl87{{\it Selected Topics in Electroweak Interactions} (Proceedings of
the Second Lake Louise Institute on New Frontiers in Particle Physics, 15--21
February, 1987), edited by J. M. Cameron \ite~(World Scientific, Singapore,
1987)}
\def \ky85{{\it Proceedings of the International Symposium on Lepton and
Photon Interactions at High Energy,} Kyoto, Aug.~19-24, 1985, edited by M.
Konuma and K. Takahashi (Kyoto Univ., Kyoto, 1985)}
\def \mpla#1#2#3{Mod.~Phys.~Lett.~A {\bf#1} (#3) #2}
\def \nc#1#2#3{Nuovo Cim.~{\bf#1} (#3) #2}
\def \np#1#2#3{Nucl.~Phys.~{\bf#1} (#3) #2}
\def \pisma#1#2#3#4{Pis'ma Zh.~Eksp.~Teor.~Fiz.~{\bf#1} (#3) #2[JETP Lett.
{\bf#1} (#3) #4]}
\def \pl#1#2#3{Phys.~Lett.~{\bf#1} (#3) #2}
\def \plb#1#2#3{Phys.~Lett.~B {\bf#1} (#3) #2}
\def \pr#1#2#3{Phys.~Rev.~{\bf#1} (#3) #2}
\def \pra#1#2#3{Phys.~Rev.~A {\bf#1} (#3) #2}
\def \prd#1#2#3{Phys.~Rev.~D {\bf#1} (#3) #2}
\def \prl#1#2#3{Phys.~Rev.~Lett.~{\bf#1} (#3) #2}
\def \prp#1#2#3{Phys.~Rep.~{\bf#1} (#3) #2}
\def \ptp#1#2#3{Prog.~Theor.~Phys.~{\bf#1} (#3) #2}
\def \rmp#1#2#3{Rev.~Mod.~Phys.~{\bf#1} (#3) #2}
\def \rp#1{~~~~~\ldots\ldots{\rm rp~}{#1}~~~~~}
\def \si90{25th International Conference on High Energy Physics, Singapore,
Aug. 2-8, 1990}
\def \slc87{{\it Proceedings of the Salt Lake City Meeting} (Division of
Particles and Fields, American Physical Society, Salt Lake City, Utah, 1987),
ed.~by C. DeTar and J. S. Ball (World Scientific, Singapore, 1987)}
\def \slac89{{\it Proceedings of the XIVth International Symposium on
Lepton and Photon Interactions,} Stanford, California, 1989, edited by M.
Riordan (World Scientific, Singapore, 1990)}
\def \smass82{{\it Proceedings of the 1982 DPF Summer Study on Elementary
Particle Physics and Future Facilities}, Snowmass, Colorado, edited by R.
Donaldson, R. Gustafson, and F. Paige (World Scientific, Singapore, 1982)}
\def \smass90{{\it Research Directions for the Decade} (Proceedings of the
1990 Summer Study on High Energy Physics, June 25 -- July 13, Snowmass,
Colorado), edited by E. L. Berger (World Scientific, Singapore, 1992)}
\def \stone{{\it B Decays}, edited by S. Stone (World Scientific, Singapore,
1994)}
\def \tasi90{{\it Testing the Standard Model} (Proceedings of the 1990
Theoretical Advanced Study Institute in Elementary Particle Physics, Boulder,
Colorado, 3--27 June, 1990), edited by M. Cveti\v{c} and P. Langacker
(World Scientific, Singapore, 1991)}
\def \yaf#1#2#3#4{Yad.~Fiz.~{\bf#1} (#3) #2 [Sov.~J.~Nucl.~Phys.~{\bf #1} (#3)
#4]}
\def \zhetf#1#2#3#4#5#6{Zh.~Eksp.~Teor.~Fiz.~{\bf #1} (#3) #2 [Sov.~Phys. -
JETP {\bf #4} (#6) #5]}
\def \zpc#1#2#3{Zeit.~Phys.~C {\bf#1} (#3) #2}

\end{document}